# Prometheus Induced Vorticity In Saturn's F Ring

Phil J. Sutton

p.j.sutton@lboro.ac.uk

+44(0)1509 228401

Feo V. Kusmartsev

f.kusmartsev@lboro.ac.uk

+44(0)1509 223316

Astronomy Unit, Physics department, Loughborough University, Loughborough, Leicestershire, LE11 3TU, UK

Corresponding author: Phil Sutton, p.j.sutton@lboro.ac.uk

# Abstract

Saturn's rings are known to show remarkable real time variability in their structure. Many of which can be associated to interactions with nearby moons and moonlets. Possibly the most interesting and dynamic place in the rings, probably in the whole Solar System, is the F ring. A highly disrupted ring with large asymmetries both radially and azimuthally. Numerically non-zero components to the curl of the velocity vector field (vorticity) in the perturbed area of the F ring post encounter are witnessed, significantly above the background vorticity. Within the perturbed area rich distributions of local rotations is seen located in and around the channel edges. The gravitational scattering of ring particles during the encounter causes a significant elevated curl of the vector field above the background F ring vorticity for the first 1 – 3 orbital periods post encounter. After 3 orbital periods vorticity reverts quite quickly to near background levels. This new found dynamical vortex life of the ring will be of great interest to planet and planetesimals in proto-planetary disks where vortices and turbulence are suspected of having a significant role in their formation and migrations.

Additionally, it is found that the immediate channel edges created by the close passage of Prometheus actually show high radial dispersions in the order $\sim 20 - 50 \; cm/s$, up to a maximum of $1 \; m/s$. This is much greater than the value required by Toomre for a disk to be unstable to the growth of axisymmetric oscillations. However, an area a few hundred km away from the edge shows a more promising location for the growth of coherent objects.

Key words: planets and satellites: formation, planets and satellites: rings, methods: numerical, turbulence, planet–disc interactions, protoplanetary discs

## 1. Introduction

Saturn's F ring comprises of many components from a dense central core that radially lies at a very narrow paradoxically stable region [1]. Here, the central core is found to be most stable very close to where traditional Lindblad resonances (spiral density waves) are generally destabilising. Straddling the central core reside less dense spiral strands formed through direct collisions with core crossing moonlets [2,3]. The central core is known to house significant populations of small moonlets, many of which were discovered from stellar occultation's [4, 5, 6], the occurrence of mini jets [7, 8] and fan structures originating out from the core (Beurle et al 2010). The connection between Prometheus and the F ring is generally well understood with many of the immediate structures and the large population of transient moonlets directly attributed to Prometheus [2, 9, 10, 11, 12, 13]. An important question still without an answer was raised by recent work into the locations of these small moonlets responsible for creating mini jets in the central core. There does not appear at least first order, to be a connection with the locations of mini jets (central core moonlets) and the encounters of Prometheus on the F ring [7]. These moonlets are typically smaller with eccentricities that are more closely matched to the central core than the moonlets known to create spiral strands [2, 14]. Is there additional dynamics at play in the F ring – Prometheus system that might help explain a seemingly chaotic distribution of central core moonlets?

## 2. Method

The numerical method used was the same as employed in our previous work of the F ring [9, 10]. Thus, the initial conditions and integration methods of numerical models can be found within. We used the code GADGET-2 [15] to model the F ring with a central core, inner and outer strands. A total of $6.5 \times 10^5$ particles were used for the simulation. GADGET-2 assumes collisionless dynamics and does this by reducing gravitational forces within a set smoothing length. For particles with mass the smoothing length was set to values just above their physical size when assuming an internal density of ice 0.934g/cm³. The smoothing length was also well within the Hill radii of particles which is given by,

$$R_{Hill} \sim a \sqrt[3]{\frac{m}{3M}} \quad [1]$$

For ring particles with mass $m$ around Saturn ($M$) with semi-major axis $a$. Although the GADGET-2 code has the ability to model gas particles we only use gravitationally interacting particles with no additional hydrodynamical forces applied during their evolution. This keeps in line with the non-gaseous planetary ring known to exist around Saturn.

To better probe the dynamical F ring system and study the idea that turbulence might play a role the evolutionary dynamics of F ring particles post Prometheus encounter we created a method to find the curl of the velocity vector field, or vorticity in the F ring. Vorticity is the tendency of a fluid to rotate where non-zero components are present it could be sign of turbulence in a fluid. Two dimensionally we can therefore show that the curl of the velocity vector field for F ring particles is:

$$\nabla \times V = \frac{\partial V_y}{\partial x} - \frac{\partial V_x}{\partial y}. \quad [2]$$

Where, $\partial V_y/\partial x, \partial V_x/\partial y$ are partial derivatives of the velocity field and $x, y$ positions of the same particle in different snapshots. The curl is well defined for each particle and changes with time. It is defined as $\nabla \times V = \frac{\delta V_y}{\delta x} - \frac{\delta V_x}{\delta y}$, where $\delta Vy, \delta Vx$ are the velocity change of the single particle and $\delta x, \delta y$ are the positions change of the same particle in different consequent snapshots. Taking regular time snapshots we track such changes and the local vorticity associated with each particle.

However, to drastically decrease the overall computational cost we opted to only calculate vorticity for a localised area around the Prometheus encounter and previously created streamer-channels as the system evolves. Where the number of calculations done is $\propto N^2$ a significant improvement in the vorticity calculations was achieved. Programs written in IDL were used to find the curl of particles velocity vectors time dependently. Due to the orbital motion of ring particles around Saturn there will always be an inherent non-zero curl present. For the F ring this is of the order $10^{-4}\ rad.s^{-1}$. Our analysis looked for changes away from this background vorticity during and after the close passage of Prometheus at minimum separation. A value of ~$1.23 \times 10^{-4}\ rad.s^{-1}$ is the background vorticity associated with the centre of the F ring.

To spatially investigate some dynamical changes in particles trajectories during and after the close encounter we created surface rendered maps depicting radial velocity dispersion. Here, we used radial velocities of particles that deviated away from the Root Mean Squared (RMS) of the F ring radial velocity. This allows an important spatial investigation of how radial dispersions change during the encounter and how they relate to other quantities like density and vorticity. All visualisation of the GADGET-2 data is done with SPLASH [16]

## 3.     Vorticity of the particles motion

When calculating the curl of the velocity vector field we do not place the F ring in any kind of reference frame, instead looking at the unaltered positions and velocities of particles. As we are only considering a two-dimensional flow the resultant curl of the velocity vectors is parallel to the z axis and of a scalar quantity. Firstly, we should address the need to probe the F ring's vorticity. Vorticity in larger protoplanetary disks has numerically been shown to 1) solve the short fall in timescales required to form 1km planetesimals from the standard accretion theory, 2) vastly increase migration rates for planets forming in the disk.

What we find in the F ring is that most of the particle's vorticity that deviates away from the normal background vorticity ($1.23 \times 10^{-4}\ rad.s^{-1}$) shows local rotations in the flow that would represent ~ 10 – 100 times per orbital period at their maximum.  Effectively this is then 10 – 100 times greater than the background value. We find that with the calculation for the curl of the vector field most of the non-zero (in this case deviations away from the background vorticity of the F ring), occurs in and around the gravitationally disrupted area. All particles with elevated vorticities are seen downstream of the Prometheus encounters in the F ring.

When channels are at their most open (typically when Prometheus is at apoapsis) radial velocity dispersions have been found to be at their lowest [11]. The radial dispersions are then also seen to increase as the channel fills back up. If the vorticity we are measuring is a result or artefact of these known phenomena then we would also expect some similar variation in vorticity throughout the orbital period of Prometheus. Along with a minimum vorticity when channels are at their most open. However, when the channels are at their most open we find an elevated non-zero component to the curl of the velocity vector field. Typically this occurs on the channel edges, with the highest non-zero values being located on the edge facing away from Prometheus. This can be seen in Fig 1 below. The vorticity does not appear to show any signs of being intrinsically linked to the radial dispersion of particles during the orbital period of Prometheus.

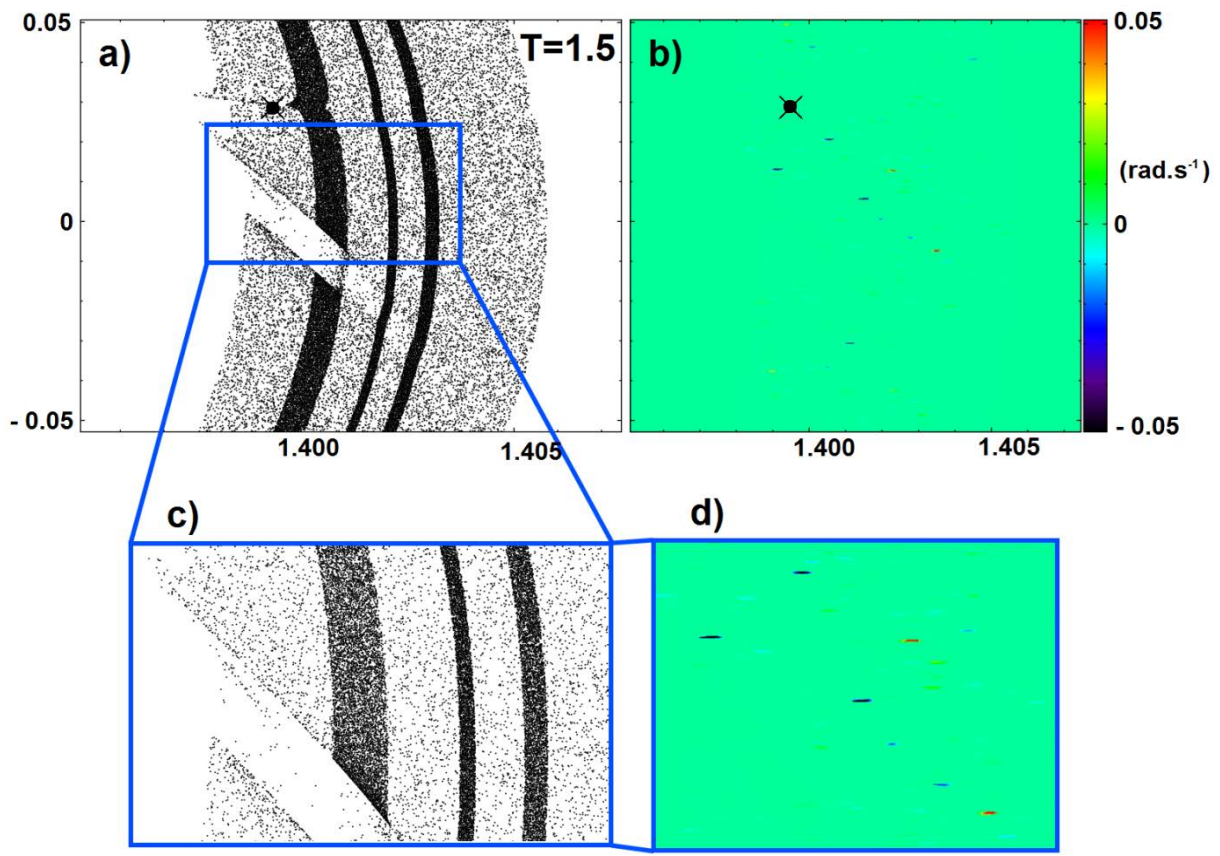

**Figure 1 |** A rendered vorticity map taken at a time T=1.5 Prometheus orbital periods. This corresponds to the first instance that the channel formation is at its most open. Both the y & x axes scales are $10^5$ km. Prometheus is marked as the circle with cross through it where it is visible in the plots. A true 1:1 aspect ratio of the F ring is not used instead it is stretched in the radial direction to

make the structures easier to visualise. The elevated vorticity can be seen to predominately occur downstream from the encounter, mostly above the channel edge facing away from Prometheus. Also note that the background Keplerian vorticity of the F ring particles is not visible in the limits we have used for our rendered plot as it is many magnitudes smaller than the maximum / minimum vorticity witnessed for those particles involved into the Prometheus encounter.

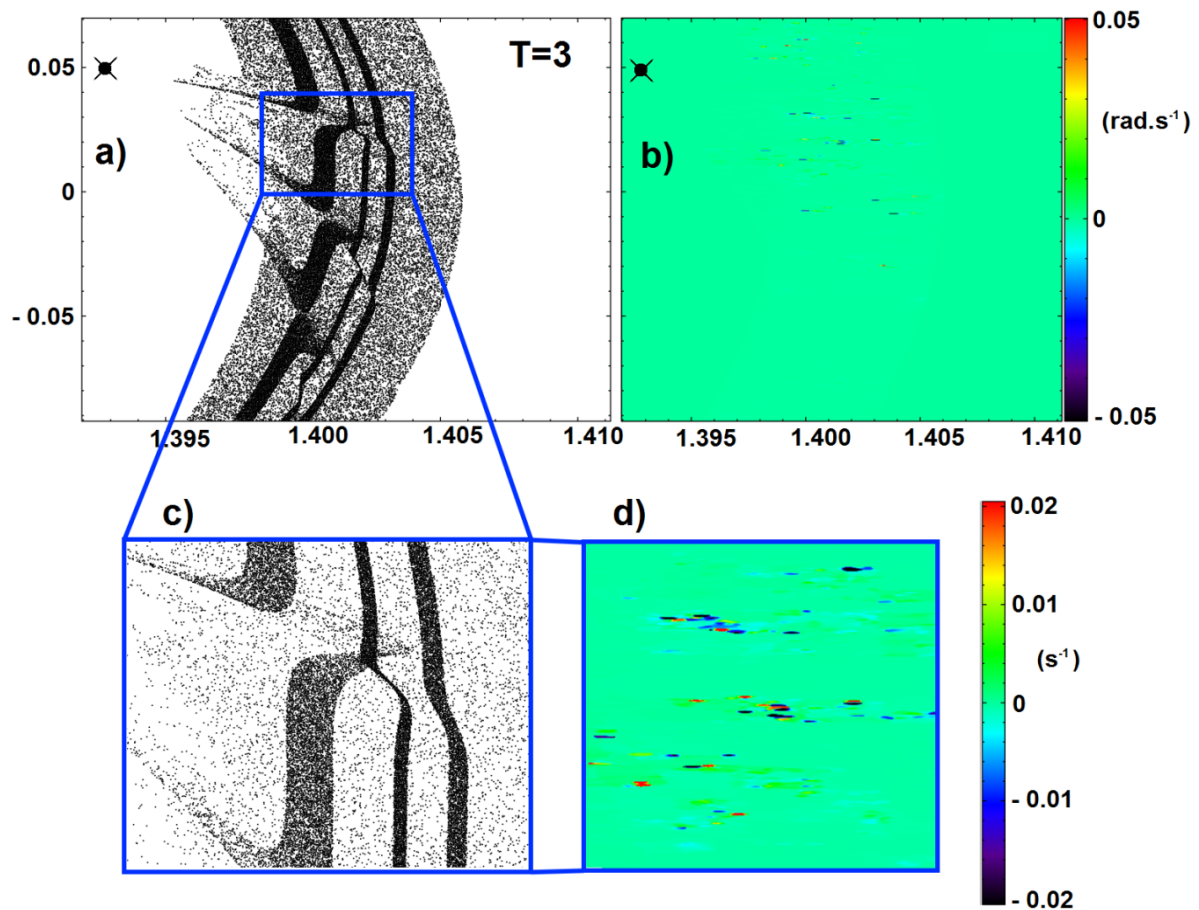

**Figure 2 |** A rendered vorticity map taken at a time T=3 Prometheus orbital periods. Both the y & x axes scales are $10^5$ km. Prometheus is marked as the circle with cross through it where it is visible in the plots. A true 1:1 aspect ratio of the F ring is not used instead it is stretched in the radial direction to make the structures easier to visualise. At this time 3 streamers are seen at their most radially inward positions. A clustering of vorticity occurs around the inner strand and central core, with two distinct areas visible in the zoomed frames (c + d). These two areas of clustering are the location that will, in 0.5 orbital phase later, form the edges of the channels. Also note that the background vorticity of the F ring is not visible on the scale we have used for our rendered plot as it is many magnitudes smaller than the maximum / minimum vorticity observed and the Prometheus encounters.

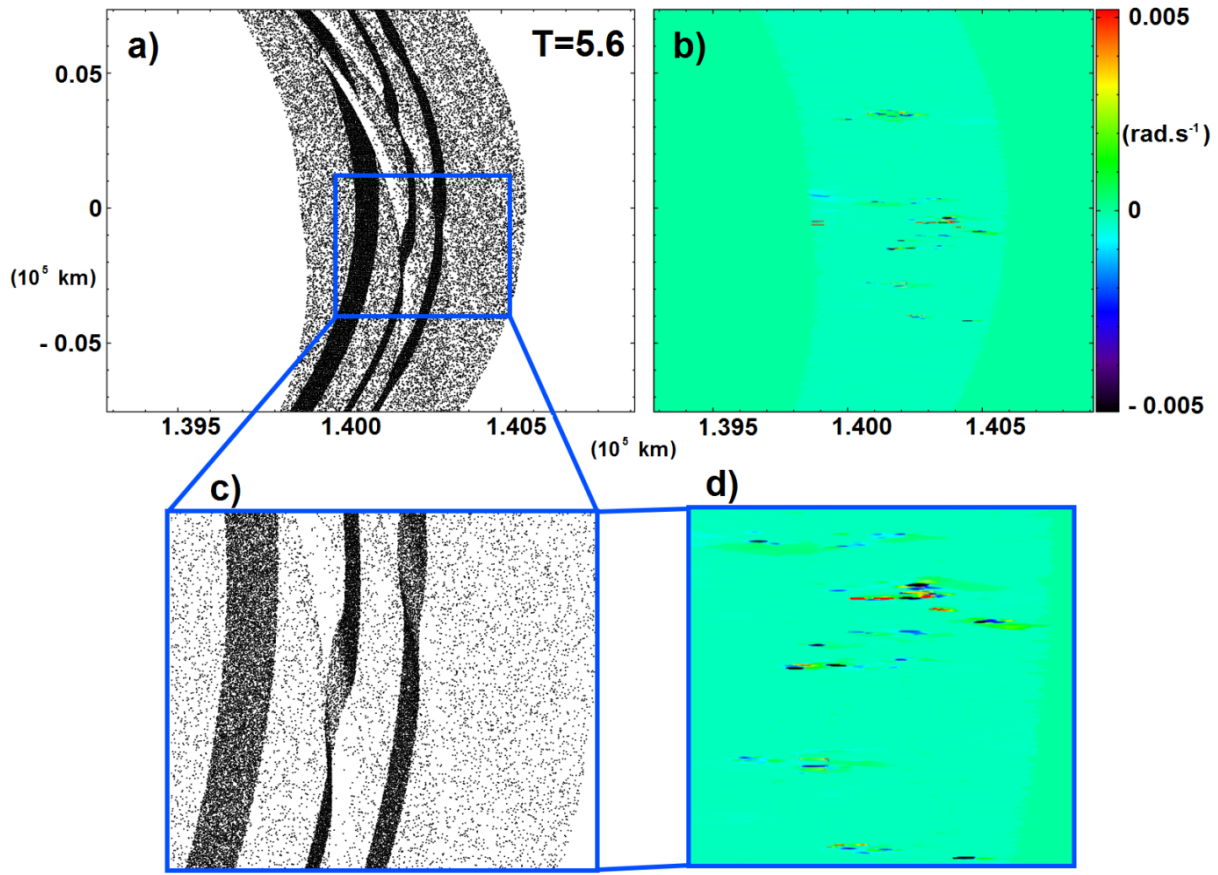

**Figure 3 |** A rendered vorticity map taken at a time T=5.6 Prometheus orbital periods. Both the y & x axes scales are 10⁵ km. A true 1:1 aspect ratio of the F ring is not used instead it is stretched in the radial direction to make the structures easier to visualise. At this time the channels are just starting to fill back up by perturbed particles after their most open phase. The channels that are 4 and 5 orbital periods old are centred in the frame with the zoomed frames c) and d) concentrating on the furthest out regions of these channels. A clustering of vorticity occurs around the inner strand and central core channel edges. The peaks of vorticity in these clusters arise exactly in the areas around the channel edges.

As the streamers are formed and then evolve on their channel formation phase we see two distinct areas of clustering in the vorticity. This can be seen in Fig 2 (c &d) where the two detached ends of the inner strand will become opposing sides of the channel 0.5 orbital phase later. Again it is this area of strong turbulence that also shows the locations of the highest densities. After the most open phase of the channel particles move back into the void created by the channel. The clustering of vorticity around the channels and edges is still clear to be seen, Fig 3. These intense particles rotations described by their velocity vector field could have an effect on any already embedded moonlets or loosely bound clumps at these locations preventing their growth or vice versa creating more clumps.

## 4. Radial Velocity Dispersions

Another important factor to consider when looking at disk stability and the growth of clumps or coherent objects through self-gravity is radial velocity dispersions. Here we make comparisons to the way we performed our changes in velocity, [10]. The deviations away from the expected Keplerian velocity magnitudes of ring particles provide us with immense data about microscopic dynamics of Saturn's ring, having a large variety of different regimes. Therefore, in addition we created a new analysis tool to spatially probe radial velocity dispersion as is generally assumed for ring / disk dynamics. Here, we first convert our Cartesian velocity vectors of individual particles into their Polar coordinates where position vectors change to $\varphi_i$ and $r_i$ and their velocity vectors in tangential and radial directions to the flow, that is, $V(\varphi)$ and $V(r)$, respectively. For each particle we now take the Root Mean Squared (RMS) of their F ring radial velocity component. The set of all deviations (or Variations) from this RMS value calculated for each individual particle comprise the radial velocity dispersion. This is then spatially rendered to create a map in the same manner as our velocity magnitude deviation maps, which are also shown in Fig 4b as a comparison. Our results are then presented in Fig 4 & 5, which are compared with existing literature. The spatial analysis of radial velocity dispersion shows two distinct elements worth noting. First of all we note that the highest radial dispersion is seen in the channel itself and is separated into two main groups. One group, which is situated at a smaller radial position (see, the left hand side of the channel in Fig 4c), maybe identified to the previously reported island of particles formed inside the channel [12]. This happens when the alignment of Prometheus' elliptical orbit and that of the eccentric F ring are at anti-alignment and thus Prometheus makes its closest approach to the F ring. At this ring-Prometheus configuration an island of particles has been created inside the channel. Note that at this moment the channels are at their most open phase. Our radial velocity dispersions show that particles in this island are collectively moving radially very fast out of the channel (speeds of > 8 m/s). This is also true for a similar group of particles arising at the opposite end of the channel (at larger radial locations around the inner strand). The variation in radial velocity dispersion arising here can be explained by a difference in orbital stage or phase of the perturbed F ring particles. In its simplest terms ring particles at the channel edges are at a different stage of their orbits than those inside the channels. These two distinct areas can be identified in Fig 4c. Secondly, the maximal amplitudes of the radial dispersion at the channel edges are considerably higher than previously reported (Fig. 5) and radial dispersion speeds of +50 cm/s are wide spread across the channel edges.

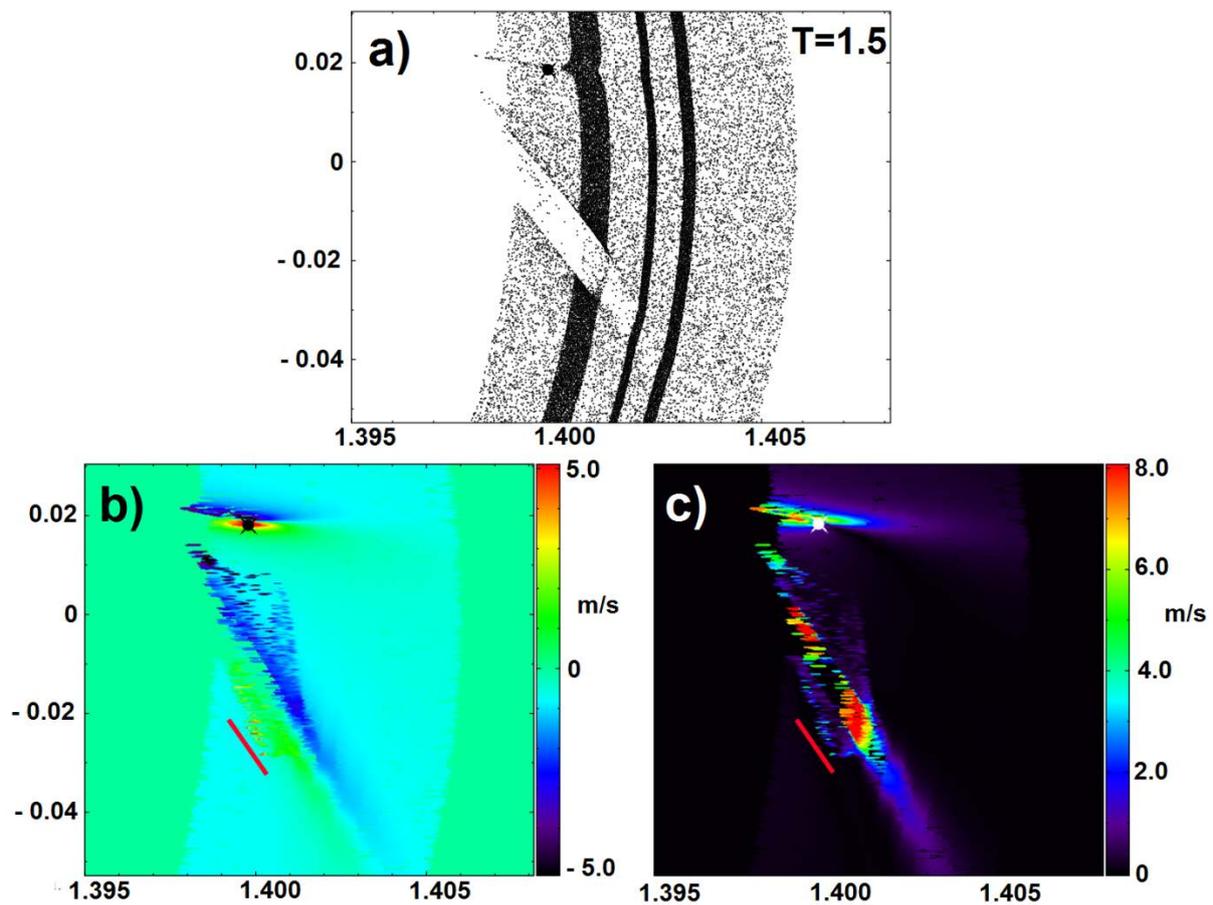

**Figure 4 |** Taken at a time when the first channel is at its most open phase T=1.5 Prometheus orbital periods (also coincides with the closet approach of Prometheus) we compare the rendered maps of our original velocity deviations (b) with the more commonly used radial velocity dispersion (c). Both the X and Y axis are scaled to $10^5$km. Prometheus is marked as the circle with cross through it where it is visible in the plots. A true 1:1 aspect ratio of the F ring is not used instead it is stretched in the

radial direction to make the structures easier to visualise. The red line is given as a guide for the eyes for identification of the same position on the ring presented in different plots.

Previously reported by Beurle et al (2010) [11] radial dispersion showed their lowest amplitude and minimum values when channels were at their most open phase. This concedes with the enhanced density on the channel edges and also with the locations of embedded moonlets [11]. Radial dispersions were found to be $< 2 cms^{-1}$ , which satisfied the Toomre parameter for the F ring. Thus, it was deemed suitable for clumps at these locations to be able to grow through self-gravity.

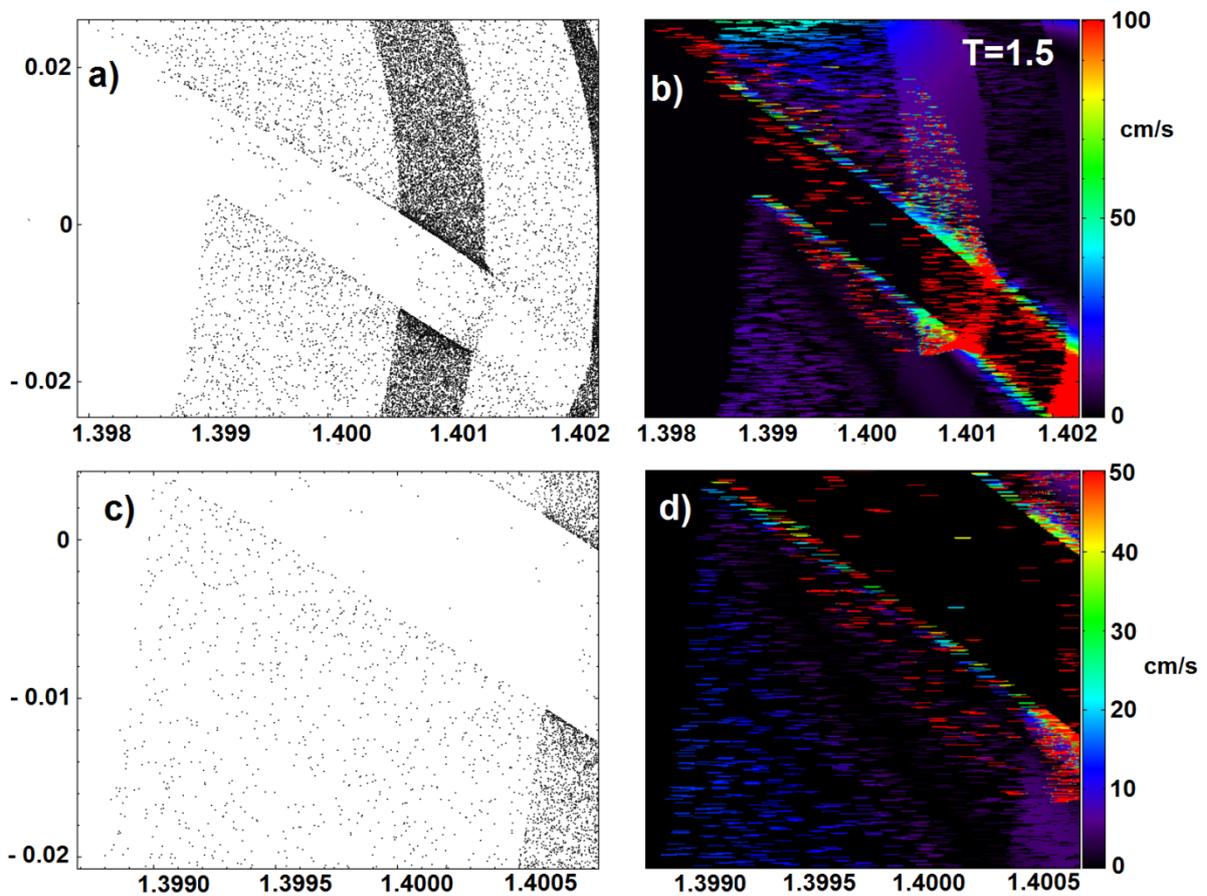

**Figure 5 |** The first channel created after the initial encounter at 1.5 Prometheus orbital periods, zoomed in further from Fig 4. Both the y & x axes scales are $10^5$ km. A true 1:1 aspect ratio of the F ring is not used instead it is stretched in the radial direction to make the structures easier to visualise. Radial velocity dispersions of particles is rendered in b) & d), with the scale adjusted to show more clearly the radial dispersions at the channel edges.

On the other hand our models indicate that particles localised along the channel edges also show that the immediate edge is populated by particles with radial velocity dispersions of over 50 cm/s. Both the channel edges show radial dispersions up to a maximum of 1 m/s. Taken at the first fully open channel since the encounter, Fig 5, we see that the edges show significant dispersions compared with ring particles located immediately above and below the channel. In fact the area with the lowest radial dispersions can be seen just below the lower channel edge (centred in Fig 5d). Here, radial dispersions are small, less than < $0.02 m\ s^{-1}$. This value is below the critical one required by the Toomre parameter for self-gravity instabilities to occur.

According to Toomre [17] a differentially rotating self-gravitating disk can be locally unstable to axisymmetric disruptions when radial velocity dispersions $\sigma$ fall below a critical value $C_{cr}$. For the Keplerian case this is shown as

$$C_{cr} = \frac{3.36 G \sigma}{\Omega} \qquad [2]$$

For the F ring with very generous 200g/cm² surface density this would equate to critical radial velocity dispersions of $C_{cr} = 0.38$ cms$^{-1}$. This considers an F ring that isn't disrupted by any external moons. However, the close passage of Prometheus alone can create radial velocities many magnitudes of order greater than this critical value. Thus, even at the channel edges where radial velocity dispersions are at their lowest we witness that their values take an order of magnitude greater than the critical one. Although, for wake structures to form the times scales needed are much greater than density enhancements made the perturbations created by Prometheus.

When minimum dispersions are discussed by Beurle et al 2010 [11] in line with density enhancements it should be noted that approximate normalised resolution (120km x 0.5km) might have allowed for particles with the highest number densities to not have the lowest associated dispersions. I.e. with lower resolution of spatial features it's possible that the minimum radial velocity dispersions are not directly on the channel edges as reported but further into the ring away from the channel edges. This would then fit with what we have found in our simulations. In Fig 5 d we can see an area of the ring that has the lowest radial dispersions, comparable to Beurle et 2010 in magnitude. One key element is that instead of the lowest radial dispersions being located on the channel edges as previously reported we find that it is an area below the bottom channel edge in Fig 5 d.  This is away from the particles with the highest density enhancements. It should also be noted that this same feature (a pocket of particles with very low radial dispersions) is not mirrored on the other channel edge (upper channel edge Fig 5 d).

Figure 6 shows the radial dispersions of all ring particles in and round the first channel seen in Fig 5. The largest dispersions are associated with the island inside the channel and the back of the channel previously mentioned. However, we can see that there are still significant dispersions away from these areas. It should be noted that not all particles with elevated radial dispersions are associated with the immediate channel edges.

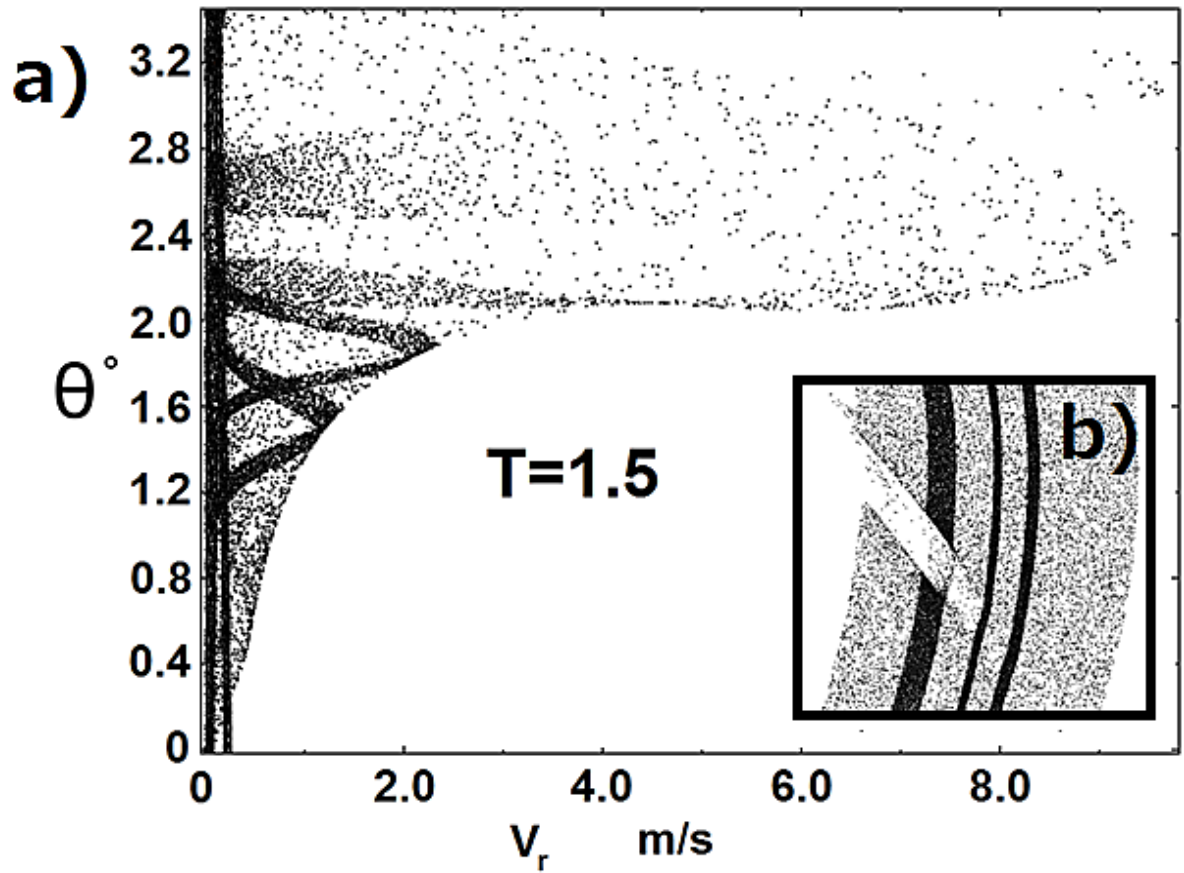

**Figure 6 |** a) Taken at the same time as Fig 4 & 5 where the y-axis represents the longitudinal position in degrees across the first fully open channel and the x-axis is the unaltered radial dispersion of each ring particle. T=1.5 Prometheus orbital periods. b) Shows the area of the channel that is plotted in (a) with the y & x axes representing longitudinal position and radial position respectively.

Now it should be noted that Fig 6 was taken at the first fully open channel. Density enhancements at the channel edges were seen to increase with each subsequent channel opening. Therefore we consider in Fig.7 the radial velocity dispersions at later times to see how it changes. We concentrate on the section of channel that is located from the inner to the central core. Due to Keplerian shear the particles with high radial dispersions become more spread out. However, both channel edges still show significant radial dispersions that are many magnitudes greater than the critical value. We see dispersive effects on the tail ends of two younger channels in the rendered frame, Fig 7b, to the right of the fully open channel. The ends of these channels are not open instead showing a large population of particles but with high radial dispersions associated with them. This is due to the larger radial locations of these sections of the channels and gravitational effect of Prometheus' close encounter. However, this is not where the highest density enhancements are seen either. Again, the same area below the channel edge facing Prometheus shows the lowest radial dispersions. At this orbital phase and time, equal to T=5.5 orbital periods or 4 Prometheus intrusions after the initial encounter this area of low dispersions spreads around $\sim 1000 km$ from the channel edge.

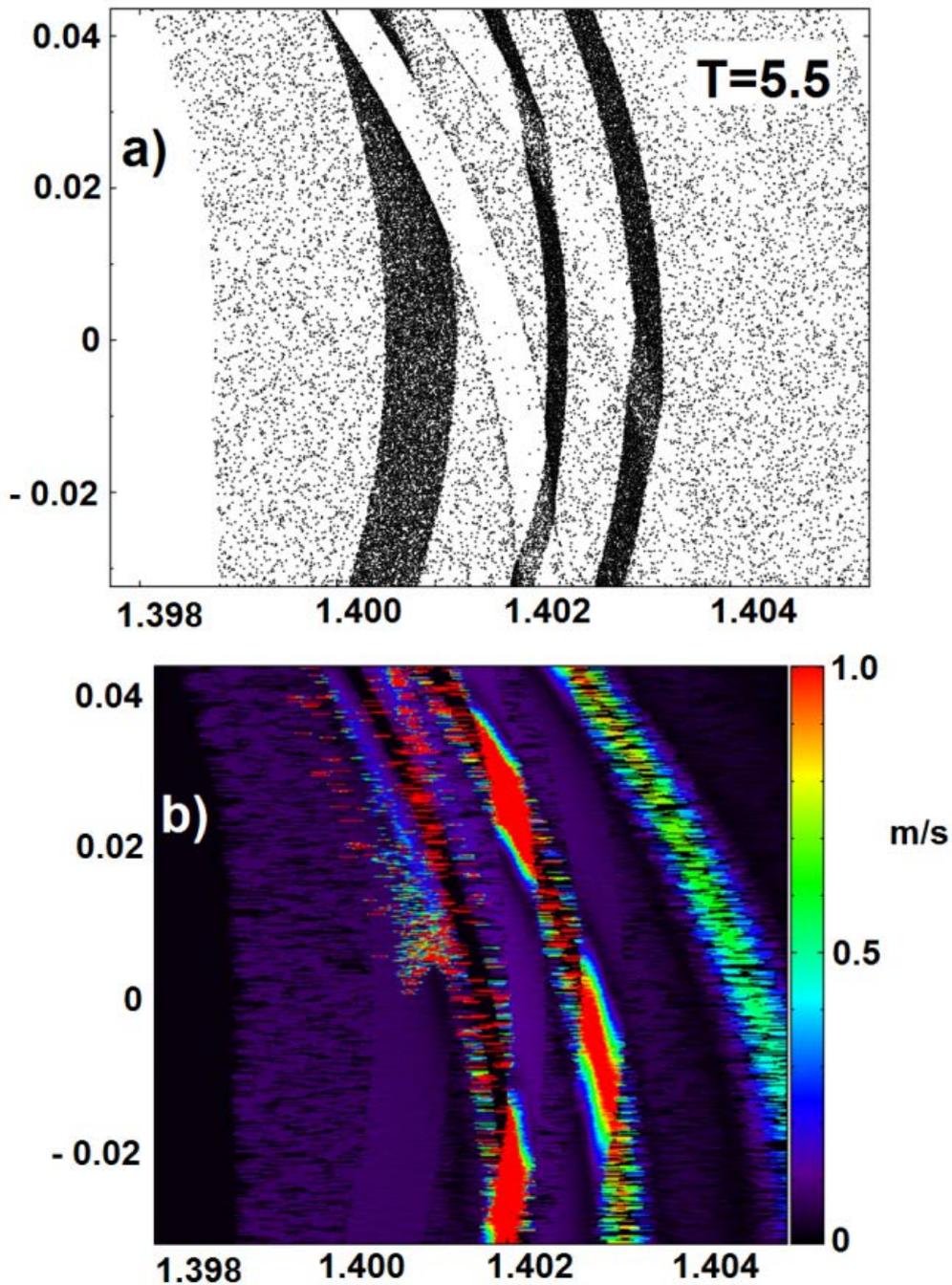

**Figure 7 |** Zoomed in look at the channel created 5 orbital periods after the initial encounter. Both the y & x axes scales are $10^5$ km. A true 1:1 aspect ratio of the F ring is not used instead it is stretched in the radial direction to make the structures easier to visualise. Radial velocity dispersions of particles is rendered in b), with the scale adjusted to show more clearly the radial dispersions at the channel edges. All of the particles within the channels exhibit radial dispersions of $> 1\, m.s^{-1}$. The oldest channel is shown to the left of the frame (5 orbital periods). Visually there is evidence of only two channels. However, a third is seen in the radial dispersions to the upper right. This corresponds to the less gravitationally interrupted part of the encounter. Although not creating a visual disruption in the ring the change in ring dynamics is still witnessed.

Furthermore we also did some supplementary investigations of dispersions perpendicular radial dispersions (azimuthal). We used the same technique as when calculating the radial dispersion of particles when channels were fully open. The only difference is that this time we only consider the longitudinal component of each particles velocity. In Fig 8 we have plotted a slightly smaller area than was used for the radial velocity dispersions. This is due to the fact the dispersions in this direction are not as significant as their radial counterparts and we centre on only the channel edges. It should be noted that due to the shearing flow in the F ring there is azimuthal dispersion of the order of meters with respect to radial positioning. However, there is one feature that is in common with dispersions in the radial direction. This is identified with the red circle in Fig 8 as being at the channel edge with a group of particles exhibiting azimuthal dispersions ~+0.5m/s more than the surrounding ring (ring particles with a comparable longitudinal positon). While significantly less than the radial dispersion observed at the same channel edge (Fig 5 b & d) which was +1m/s the azimuthal dispersion plays an important role in clump formation in such a strong tidal environment [18]. Here it is likely that ~0.5m/s azimuthal dispersions observed would be very destructive towards already formed clumps that were subject to a passing of Prometheus. It was found that collisions in strong tidal environments between particles were more dependent on collision angle with respect to orbital motion. Radial collisions were found to be the least destructive while longitudinal collisions the most destructive. Thus, further investigations into azimuthal dispersion would be more relevant to Saturn's F ring than within a circumstellar disk which can be considered more like free space.

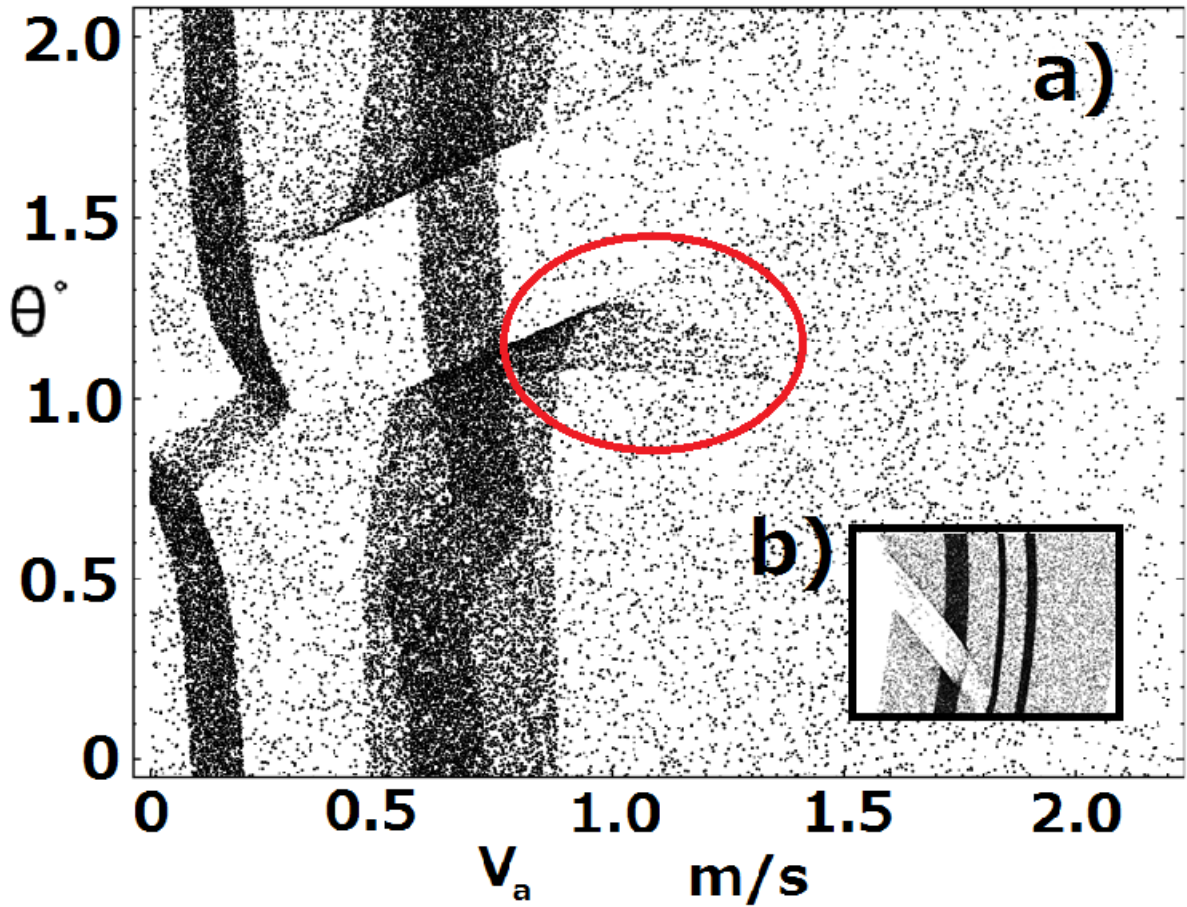

Figure 8 | a) Taken at the same time as Fig 4, 5 & 6 where the y-axis represents the longitudinal position in degrees across the first fully open channel and the x-axis is the unaltered azimuthal dispersion of each ring particle. Taken at T=1.5 Prometheus orbital periods b) shows the area of the channel that is plotted in (a) with the y & x axes representing longitudinal position and radial position respectively.

## 5. Discussion

We have found that in a nearby and fast evolving system, such as the Saturn's rings, there arises elevated vorticity which is induced due to a perturbing moon. This is beyond any vorticity that might be associated with the ring particles orbits around Saturn, which would be of the order $10^{-4}\ rad.s^{-1}$. This may appear as an obvious outcome to the moon-ring interaction but the details provided by our model does give us a detailed glimpse of this real nearby system. The key importance of the developed models is that it well describes dynamical vortices and associated rotational flows of the ring particles with an intriguing indication towards the existence of turbulence. The physical picture which we developed originally for the dynamics of the Saturn ring is very general and can be applied to other systems such as protoplanetary disk. Dynamical vortices and turbulence we found can significantly affect how planets and planetesimals form and migrate, e. g. within a larger protoplanetary disk [19, 20, 21, 22, 23, 24, 25, 26, 27, 28, 29, 30]. Type III migration of a planet in a disk can be much faster than other types of migration and mostly related due to the planets interaction with large scale vortices within the disk. The trapping of planetesimals within cores of vortices can also aid in their growth and overcome the short fall in time scales required for planet formation within the standard core accretion model. As the dust particles get larger in a gaseous circumstellar disk a gas drag causes a loss of angular momentum. This in turn means that, in the absence of other physical processes, the disk is cleared of larger dust particles as they then fall inwards to the central star [31]. A significant problem with the standard core accretion theories as it doesn't explain the rapid formation of 1km sized planetesimals we need to account for what we see. Therefore, it could be that an analogy of the F ring system is more suited to that of a debris disk where the dust – gas ratio is geared more predominately towards larger dust particles than gas (although of a different tidal environment). With many of the larger circumstellar disks the reduced tides from the host star in comparison to Saturn would mean it is of a different tidal environment. However, the reduced tides but might play a more important role in these systems being less destructive.

The removal of dust from the disk due to the gas drag therefore severely hinders the formation of the essential building blocks of planetesimals and ultimately planets. A process that seeks to compensate for this effect has been to found to theoretically exist in hydro-dynamically turbulent protoplanetary disks. Here, vortices can act as nurseries for centimetre – kilometre sized planetesimals [21]. In their numerical models it was found that vortices possessed the ability to segregate mm to cm sized particles from the gaseous protoplanetary disk, coalescing particles towards their centres where gravitational collapse helped in the rapid growth of particles to km size. The vortex capture of particles was also seen to transpire at locations appropriate to planet formation, again suggesting it to be a relevant process [28]. Clump formation from particles being drawn towards the centre of vortices was seen to happen on very short time periods, on the order $10^1 - 10^2$ orbital periods. The smallest particles were seen to settle towards the centre (the most stable points in the vortex) while the larger particles relax into orbits about the vortex edge. The stopping times of smaller particles are naturally shorter than that of larger particles due to the additional inertia they carry. The smallest particles ended up following the gas flow within the vortex. Thus, if a perturbing planet created a similar curl in the velocity vector field of a gaseous circumstellar disk, acceleration in the growth of clumps could be witnessed and directly attributed to the larger planet.

Within our own models we generally see non-zero vorticities clustered around the centre of large scale disturbances previously identified [9, 10], with quite an active non-zero vorticity along channel edges. Further investigations of the vorticity calculation shows that most of the non-zero vorticity occurs during the first 1 – 3 orbital periods after the initial encounter. Vorticity then quickly reverts close to background levels by around 10 orbital periods. This could suggest the curl of the velocity field quite quickly reduces back to pre-encounter levels. If the vorticity we calculate was associated with particles moving on out of phase elliptical orbits we would expect to still see an appropriate vorticity. The fact that we still see significant radial dispersions and velocity deviations at 10 orbital periods suggest that this isn't an artefact creating the vorticity values. Although care should be taken as the F ring isn't a true fluid.

Something else of interest to note that may be applicable to the F ring but is not tested in our models is particle spin. If rotations in the flow of particles are seen (vorticity) then particles with a physical size would likely receive a rotation themselves during the passage and subsequent gravitational scattering event of Prometheus. This additional spin of particles might have some additional important influence on the collisions of particles and the ability for some of the larger particles to withstand the already destructive tidal forces present. Experimentally the spin state of ring particles can be found by investigating the thermal emission of the rings [32, 33]. A similar investigation of the F ring with the Infrared Spectrometer (IRS) on board *Cassini* might prove useful for future understanding of the F ring. The thermal inertia of particles can help us probe the populations of the fast and slow rotators which coupled with already know particle sizes derived through stellar occultation data [34] would help experimentally probe the curling of the velocity vector field in the F ring by Prometheus.

Ultimately we find that areas of elevated velocity vector curl and some of the changes in local dynamics (velocity and accelerations) occur where embedded moonlets are found [11] and density enhancements witnessed (channel edges). Their location on channel edges being random and coupled with the elongation of the channel formations due to Keplerian shear this could result in coherent objects being chaotically positioned within the central core. These then points towards the known randomly distributed moonlets found in the central core by Cassini [7, 8] and could offer some explanation as their origin.

Investigation of radial velocity dispersions at the channel edges reveals that they themselves aren't the ideal locations for the rapid growth of clumps or moonlets. Here, dispersions (both radial and azimuthal) are high enough to hinder the natural self-gravitational collapse. Along with many of the previously discussed dynamics at the channel edges [9, 10, 11] there is also evidence that points to a difference between the two edges with regards to the radial velocity dispersions. The area just away from the edge facing Prometheus shows the lowest dispersions and is approximately 1000km away from the edge (this can be seen on the channel structure presented in Fig 5. The low dispersion area is just under the bottom channel edge). This is indeed the edge that embedded moonlets have been observed creating fan structures within the central core. The edge areas also corresponds to radial velocity dispersions of $\gg 2 cm\ s^{-1}$. These numbers are close to the critical values stated by

Toomre for a disc to be unstable to gravitational collapse. This result is in some disagreement with the proposal [11] that suggested the minimum radial dispersions occurred at the channel edges. While it is true minimum dispersions occur in the same boxes as their highest densities (also the channel edges) this could be dependent on the size boxes employed. There are areas a few hundred km away from the immediate edges (the direct edge of the channel) that could be responsible for the minimum dispersions, different than what was reported in Beurle et al (2010) [11]. The velocity magnitudes at the edges themselves are greater than the critical value and do not show a uniform dispersion, instead they range randomly from $5 cm\ s^{-1} - 100 cm\ s^{-1}$. Therefore, in our models when considering the spatial distribution of the velocity radial dispersions, it appears that the growth of coherent objects might be better suited further away from the immediate channel edges (further into the F ring away from the channel edges)

The increased number densities of particles and high radial velocity dispersions at channel edges would lead to an increase in collision rates between particles. Although the collision rates have not been estimated within our models but seems a likely outcome and it is reasonable to suggest that it would be counterproductive in the growth of clumps. It has been noted that density enhancements at the channel edges show considerable fluctuations between subsequent orbital periods [9, 11]. The enhanced densities and high radial dispersions at these locations thus make a very challenging environment for clumps to grow in. The formation of areas with highest radial velocity dispersions in islands at the back of the channel (at larger radial locations) when the channels were at their most open phase were previously reported [14]. The particles inside these islands display high radial dispersions in the order of m/s higher than in other areas. They are predominantly moving in a radial direction away from the Keplerian flow. If considered individual particles post encounter it could be that the particles at those locations are in a different phase of their new more eccentric orbits compared with the rest of the particles in the channel formations.

Obviously there are issues surrounding the very different tidal environments of the planetary rings (Saturn's F ring being discussed here) and various circumstellar disks. Some additional work ideally should be done to investigate how the local velocity vector field is distorted due to a close passage of a planet or planetesimal in a circumstellar disk. The weaker tidal environment in these systems is likely to yield different results but it might be of great interest to investigate the plausibility of young planetesimals in disks to initiate vorticities (beyond the background value of the disk). As these vorticities and associated dynamics are proposed as a potential mechanism for the rapid growth of objects within circumstellar disks it is especially of interest.

One of the main arguments for the analogy of the F ring to circumstellar disks comes from the very initial collective rotation of the particles during the close passage of a shepherding moon. The different tidal environments of, for example, planetary rings and circumstellar disks are going to have an effect on long term vorticity that would reside in the disk post encounter. However, what we report can still be useful for looking at immediate local rotations caused by a passage of nearby moon or planet to a disk. When considering Hill Radii's $\left(\sim a^3\sqrt{m/3M}\right)$ in the planetary ring the area affected by the passage is smaller than that of even a small planet embedded in a circumstellar disk. The likely immediate differences between such systems would be the scale of vorticity spatial distribution induced and along with complicated 3-D disk disruptions that would not be as important in a flat planetary disk. The different tidal environment could also be more favourable to creating a turbulent perturbation in a planetary disk than in a weaker tidal environment like in the planet forming region of a circumstellar disk. In the circumstellar disk the weaker tidal forces from the central star might mean that any vorticity induced by a planet would not be significant enough to have any impact on the acceleration in the accretion of clumps. Direct comparisons could be done by keeping the particles and central mass the same as we have investigated but instead changing the tidal environment. So, effectively move the F ring further out radially to a location that is comparable to that of a larger circumstellar disk where a planet might reside. Here, the same tidal environment can be investigated without changing any other parameters. Thus, a direct investigation of tidal environment can be done.

Additionally, considering the fairly high radial and azimuthal dispersions we have found in our work during the close passage of Prometheus would also like to note that the inclusion of collisions between particles could have an impact on the dynamics of vortices. We expect to address this issue in our next publication.

**Acknowledgements**

I would like to thank Dr Daniel Price from MONASH University who made modifications to the (SPLASH) software that meant I was able to render additional quantities from my simulations that were previously not possible. We would also like to thank Prof Heikki Salo for useful discussions which have helped improve this work.

**Author contributions**

P. J. S performed the numerical simulations, analysed the results and wrote the manuscript.

F. V. K contributed ideas, took part in discussions of results and helped with editing of the manuscript.

**Competing financial interests**

The authors declare no competing financial interests.